\documentstyle[12pt]{article}
\begin{document}
\thispagestyle{empty}
\begin{center}
\LARGE \tt \bf {On String Cosmology and de Sitter inflation with massless dilatons and dynamical torsion}
\vspace{3cm}

{\large By
L.C. Garcia de Andrade. \footnote{Depto. de F\'{\i}sica Te\'{o}rica - IF - UERJ - 
Rua S\~{a}o Francisco Xavier 524, Rio de Janeiro, RJ
CEP:20550  ,Maracan\~{a}.e-mail:garcia@dft.if.uerj.br}}
\vspace{3cm}
\begin{abstract}
Variation of the $4-D$ string cosmology action with dynamical torsion and massless dilatons lead to an expression of torsion in terms of massless dilatons in the case of de Sitter inflation.The solution is approximated according to the COBE data.
\end{abstract}
\end{center}
\newpage
Recently a renewed interest in cosmolgy with spin and torsion has arisen in the context of inflationary cosmology with the independent investigation
by Palle \cite{1} and myself \cite{2} of density fluctuations in Einstein-Cartan cosmology by making use of cosmic background radiation from the COBE data \cite{3}.Earlier also Maroto and Shapiro \cite{4} have computed the stability of string higher-order gravity cosmology de Sitter solution with dilatons.In their paper they consider a non-dynamical torsion solution where in particular a constant torsion may play the role of a cosmological constant.In this note we show that the relaxation of the constraint of a non-dynamical torsion may lead to some interesting physical consequences,such as the dependence of torsion with the massless dilaton potential.It may also contribute to a better understanding of the role played by torsion on the inflationary process.We start from a Friedmann metric
\begin{equation}
ds^{2}=dt^{2}-{a(t)}^{2}(dx^{2}+dy^{2}+dz^{2})
\label{1}
\end{equation}
where the action is given by
\begin{equation}
S=\int{dte^{3b}e^{-2{\phi}}L(T,\dot{T},\dot{\phi})}
\label{2}
\end{equation}
Where $b=log{a(t)}$.Variation of this action with respect to the dilaton field ${\phi}$ leads to the following Euler-Lagrange equation
\begin{equation}
\frac{d}{dt}(\frac{{\partial}e^{3b}e^{-2{\phi}}L}{{\partial}{\dot{\phi}}})-\frac{{\partial}L}{{\partial}{\phi}}=0
\label{3}
\end{equation}
Expansion of the first term of this last equation leads to the expression
\begin{equation}
-2{\phi}L+\frac{{\partial}L}{{\partial}{\dot{\phi}}}(3H-2{\dot{\phi}})+\frac{d}{dt}{\frac{{\partial}L}{{\partial}{\dot{\phi}}}}=0
\label{4}
\end{equation}
Where $H(t)=\frac{\dot{a(t)}}{a}$.By making use of the following Lagrangean
\begin{equation}
L=R+\dot{T}+T^{2}-{\dot{\phi}}^{2}
\label{5}
\end{equation}
which represents the Lagrangean for a 4-D string cosmology with massless dilatons and dynamical torsion.Solution of expression (\ref{5}) into expression (\ref{4}) leads to 
\begin{equation}
-6H{\dot{\phi}}-4{\dot{\phi}}^{2}-2{\ddot{\phi}}+2{\phi}[-12{H_{0}}^{2}+\dot{T}+{T}^{2}-{\dot{\phi}}^{2}]=0
\label{6}
\end{equation}
by making use of the approximation that $H^{2}<<{\dot{\phi}}^{2}$ which is consistent with the matter density fluctuation $\frac{{\delta}{\rho}}{{\rho}}=\frac{{H}^{2}}{\dot{\phi}}<<<1$ which is consistent with the COBE data, and considering that the variation of torsion is not much less than the variation of dilaton potential although torsion is indeed much weaker than the dilaton potential we obtained the following approximation consistent with the observations
\begin{equation}
T(t)={\phi}(t)+3H_{0}log{\phi}
\label{7}
\end{equation}
which reveals the behaviour of torsion in terms of a massless dilaton during the inflationary de Sitter phase.
Expansion of this expression with time reads
\begin{equation}
T(t)=a_{0}t+b_{0}t^{2}+3H_{0}{\phi}+H_{0}c_{0}{\phi}^{2}+...
\label{8}
\end{equation}
And torsion seems to act as a energy of the vacuum with massive terms ${\phi}^{2}$ although we are dealing with massless dilatons.
\section*{Acknowledgements}
We would like to express our gratitude to professors Ilya Shapiro and Rudnei Ramos for helpful discussions.Partial financial support from CNPq. is gratefully acknowledged.

\end{document}